\documentclass[conference]{IEEEtran}
\IEEEoverridecommandlockouts
\usepackage{cite}
\usepackage{amsmath,amssymb,amsfonts}
\usepackage{algorithmic}
\usepackage{graphicx}
\usepackage{textcomp}
\usepackage{comment}
\usepackage{hyperref}
\hypersetup{
    colorlinks=true,
    linkcolor=blue,
    citecolor=blue,
    filecolor=magenta,      
    urlcolor=blue,
}
\usepackage{verbatim}
\usepackage{adjustbox,lipsum}
\usepackage{xcolor}
\def\BibTeX{{\rm B\kern-.05em{\sc i\kern-.025em b}\kern-.08emT\kern-.1667em\lower.7ex\hbox{E}\kern-.125emX}}
\usepackage{comment}
\usepackage{ifthen}
\newboolean{showcomments} 
\setboolean{showcomments}{True}

\begin{document}

\title{Generating Requirements Out of Thin Air: Towards Automated Feature Identification for New Apps}



\author{{Tahira Iqbal}\\
\IEEEauthorblockA{
\textit{fortiss GmbH}\\
Munich, Germany \\
iqbal@fortiss.org}
\and
\IEEEauthorblockN{Norbert Seyff\\}
\IEEEauthorblockA{
\textit{FHNW}\\
Windisch, Switzerland \\
\textit{University of Zurich} \\
Zurich, Switzerland \\
norbert.seyff@fhnw.ch}
\and
\IEEEauthorblockN{Daniel Mendez}
\IEEEauthorblockA{
\textit{Blekinge Institute of Technology}\\
Karlskrona, Sweden \\
\textit{fortiss GmbH}\\
Munich, Germany\\
daniel.mendez@bth.se\\}
 \thanks{The research leading to these results has received funding from the European Union’s Horizon 2020 research and innovation programme under the Marie Skłodowska-Curie grant agreement No 674875.}}
\maketitle
\begin{abstract} 
App store mining has proven to be a promising technique for requirements elicitation as companies can gain valuable knowledge to maintain and evolve existing apps. However, despite first advancements in using mining techniques for requirements elicitation, little is yet known how to distill requirements for new apps based on existing (similar) solutions and how exactly practitioners would benefit from such a technique. 
In the proposed work, we focus on exploring information (e.g. app store data) provided by the crowd about existing solutions to identify key features of applications in a particular domain. We argue that these discovered features and other related influential aspects (e.g. ratings) can help practitioners(e.g. software developer) to identify potential key features for new applications. To support this argument, we first conducted an interview study with practitioners to understand the extent to which such an approach would find champions in practice. In this paper, we present the first results of our ongoing research in the context of a larger road-map. Our interview study confirms that practitioners see the need for our envisioned approach. Furthermore, we present an early conceptual solution to discuss the feasibility of our approach. However, this manuscript is also intended to foster discussions on the extent to which machine learning can and should be applied to elicit automated requirements on crowd generated data on different forums and to identify further collaborations in this endeavor.\\

%

\end{abstract}


\section{Introduction}

Requirement Engineering (RE) is crucial for successful product development\cite{ZHIJIN2018}. Requirement elicitation, in particular, is one of the major activities in RE. Often, requirements elicitation tends to be limited to face-to-face meetings, or interviews, prototyping \cite{{Zowghi2005}, {RE_roadmap}}.\\
Crowd based approaches are becoming more prominent \cite{{Crowdsourcing_in_SE},{Huang:2013},{Eduard.C}}. For example, social and mobile systems are reaching out to a vast number of highly distributed and heterogeneous stakeholders ~\cite{Crowd_RE}. Such systems often provide different means that allow their users to communicate their option about the app (e.g. feedback in app stores). Conventional requirements elicitation methods  \cite{{Zowghi2005}, {RE_roadmap}} do not address how this nowadays widely available data can be used to elicit requirements. 
However, the data generated by the crowd of users open the door for developing novel requirements approaches and is becoming a prominent source for eliciting and prioritizing requirements as well as for planning releases. This fits into a recent trend in requirement engineering research focusing on the application of data-driven approaches. Researchers \cite{{Maleej_2016},{Maleej2015},{RE2018FAME}} have already demonstrated how to use the information provided by the crowd to support developers.

It is important to note that the focus of recent research in this domain is on software evolution and not on developing new applications. For developing a new app, it is important to have knowledge about applications existing in a particular domain, including their key features and end-user responses about those features. This information can, for example, be found in app stores. There could be many existing apps ~\cite{SANER_2016} having similarities to the intended app to be developed. This information provided by the crowd e.g. app descriptions, screenshots, and user feedback can provide important input for requirement elicitation. We envision that the requirement elicitation processes for new apps can be stimulated by mapping existing similar features from other apps to the new app based on different parameters such as ratings and user reviews. We foresee that such an approach can help to reduce known elicitation problems such as time and scoping issues \cite{RE_systematicREview:2006}.

In this paper, we present early research results and outline a research roadmap to generate requirements out of the various data sources for the development of new apps. Section \ref{sec: Goal and Research Questions}, defines the overall goal of our work and discusses key research questions. In Section \ref{sec:Interview Study}, we present our interview study. Section \ref{sec:key_finding} lists key findings, whereas the next steps of our research and the initial proposed solution are discussed in Section \ref{sec:RoadMap}. In Section \ref{sec:Related_Work}, we discuss related work before concluding our paper in section \ref{sec:conclusion}.

\section{Goal and Research Questions}
\label{sec: Goal and Research Questions}
The goal of our research is to support requirement elicitation for new apps in particular domains. To achieve this goal, we investigate how existing data from similar apps can be used to identify key features. In particular, our idea is to perform an identification of relevant features based on a product vision and available response from the crowd on online forum e.g. social media, app store. This will allow us to provide a list of features from existing apps relevant to the new application. Additional selection criteria could be the popularity of features. 

The long-term vision of our overall research is to develop an automated solution supporting the mobile application development industry. Our approach could help companies to reduce requirement elicitation problems for new apps such as scoping, understanding, and volatility \cite{wahono2003analyzing}. In particular, we envision that companies applying our approach would gain valuable knowledge in the early stages of app development, reduce time, and effort for requirement elicitation \cite{RE_systematicREview:2006}. We argue that they will already have a better understanding of existing apps by analyzing the users' opinions at the feature level and by knowing the risks associated with them, e.g., possible bugs, user likeness, and acceptability. 

To achieve this long-term goal, we are following a Design Science research approach\cite{Design_Science} where we iteratively develop and constantly revise technical solution proposals based on an increased understanding of the problems. This research is steered by the following research questions:
\begin{itemize}
   
\item RQ 1: What are contemporary practices and challenges in requirements elicitation for developing new apps?
\item RQ 2: Do practitioners already analyze crowd-generated data or information provided by the crowd e.g. app store data and if so, how?
\item RQ 3: How can we link user feedback from the crowd to individual features extracted from app descriptions using machine learning?
\item RQ 4: What are the possible influential factors for suggesting features for the new app? 
\item RQ 5: How can we systematically map features from existing apps to the new app? 
\end{itemize}

RQ 1 aims at understanding the current requirement elicitation practices applied in industry for new app development. Under RQ 2, we aim at investigating the processes and associated problem with crowd provided data e.g.  app store mining processes and their associated problems. Answering these two research questions will help us to understand the needs of practitioners and the availability and acceptability of potential solutions. In RQ 3, we aim at developing an approach which allows us to identify and analyze user feedback given for a particular feature e.g. features mentioned in the app description in app stores. However, feedback from the crowd is significantly large in size, and it is hard to identify what, when, and where a user is mentioning a particular feature. RQ 4 aims at identifying different factors that can influence a feature suggestion process for a new app such as app ratings and user sentiments. Based on the factors identified in RQ 4, features have to be selected from existing apps. RQ 5 focuses on this mechanism of selecting features for the new app.

The first two research questions aim at framing the problem, and the latter ones aim at designing and evaluating a specific solution which depends on the outcomes of RQs 1 and 2. This paper presents our ongoing research and discuss preliminary results for RQ1 and RQ2. To answers RQs 3-5 explanations and ideas are provided in the scope of future work.
 
\section{Interview Study}
\label{sec:Interview Study}
To understand the current state of practice of requirement elicitation for new apps (RQ 1 and RQ 2), we conducted an exploratory study with eleven practitioners from different companies. To this end, we conducted semi-structured interviews to get in-depth insights into the interviewees' worlds, opinions, experiences, and feelings~\cite{Interview_Simula}. The detailed analysis of the study results is still ongoing, but for this paper we have identified first key findings.

\subsection{Instrument}

After designing a first prototype of our instrument, we conducted mock interviews with three people; one from industry and two from our research group. These interviews were not included in the analysis. From these interviews, we estimated the duration of an interview, and identified ambiguities in our questions. We identified possible misunderstandings and ambiguities by explaining each question to the interviewee and discuss their relevance to different scenarios. The final iteration to our questionnaire was performed on the basis of inputs from mock interviews. The interview questionnaire is online available
\cite{Xyz2019}\footnote{See also \url{https://tinyurl.com/yxegurs6}.}

Each interview was designed in three folds; the first part dealt with participant's profile related information, the second part was about their current practices for app development and associated challenges, and the last part focused on the identification of feature mapping mechanism from existing application to new application. For our study, we tried to follow the practice and design principle discussed in \cite{Interview_Simula}.

\subsection{Selection of Participants and Demographics}

For the interview, we started contacting app developers from our personal network via email. We shared the brief concept and details for our study in this email. During that email exchange, we further narrowed down the list of potential candidates base on their experience and profiles. The participants required for our study should have an overview of the requirement engineering process and experience in the mobile application development industry. For our study, we restricted ourselves to requirement engineers, business architects, project managers, consultants, and software engineers who know about the requirements engineering activities and application development process of their companies. 

The study is based on 11 participants from central Europe. In the first half of the interview, we collected profile information of the interviewee e.g. information related to their application, experience in RE and application development experience, application related. 

Out of 11 participants, only 1 had 2-4 years of experience, 5 reported 4-6 years of experience, 3 participants had 7-10 years of experience, and 2 said that they had more than 10 years of experience. We did not include anyone with less than 2 years of experience, and the average experience was almost 6 years.

We focus on the essential demographic characteristics and omit those we consider irrelevant to our study, such as age, gender, or nationality. Interviewee profile-related information is summarised in Table \ref{tab:overview}.
\begin{table*}
 \centering
\begin{tabular}{|p{0.5cm}|p{2.9cm}|p{1.4cm}|p{2.5cm}|p{2cm}|p{2.5cm}|}
\hline
\textbf{No.} & \textbf{Role} & \textbf{Year of Experience} & \textbf{Development Type} & \textbf{No. of App Downloads} & \textbf{App Domain} \\ 

\hline
1 & Designer, Requirements Engineer & 4-6 & Customer Development 
& 1.3M & Banking and Finance\\ 
\hline
2 & Requirements Engineer & 4-6 & Customer Development 
& 1K & Social\\ 
\hline
3 & Software Developer & 7-10 & Customer Development 
& 3K & Game, Productivity\\ 
\hline
4 & Project Manager, Developer & 7-10 & Product Development
& 20K & Productivity\\ 
\hline
5 & Team Leader & 7-10 & Customer Development and Product Development
& 1M & Banking and Finance \\ \hline
6 & Requirements Engineer, Project Manager & 4-6 & Customer Development
& 20K & Health and Fitness\\ 
\hline
7 & Software Developer 
& 4-6 & Customer Development and Product Development 
& 1M & Business, Finance\\ \hline
8 & Project Manager, Requirements Engineer & \textgreater10 & Customer Development and Product Development
& 40K & Productivity, Education, Lifestyle\\ \hline
9 & Software Developer &  2-4 & Product Development 
& - & Maps \& Navigation\\ 
\hline
10 & CEO & \textgreater10 & Customer Development and Product Development
& - & Multiple \\ 
\hline
11 & Project Manager, Software Developer & 4-6 & Customer Development and Product Development & - & Lifestyle (energy)  \\ \hline
\end{tabular}
\caption{Interviewee Profile Overview}
\label{tab:overview}
\end{table*}

For privacy reasons, particpant's name, company name, and the names of the applications are omitted. The majority has experienced developing mobile applications on both of the most wide-spread platforms iOS and Android. For a better understanding, we have provided the domain of application, the number of downloads and whether the app is customer or product based.

\subsection{Data collection and analysis}
The interviews were conducted over Skype with a duration of  35-55 minutes with an average of almost 45 minutes. Interviews were recorded if permission was granted from the interviewee. In all cases, the interviewees agreed on audio recording without hesitation. 
    
For the data analysis, the first author transcribed interviews and then structured information with help of notes that were taken during interviews. In the next step, techniques for transcribed  interviews analysis and extracting key findings were discussed. The potential patterns, key observations and interesting aspects related to our questions were initially analyzed and allowed us to identify the key findings presented in this paper. We are currently conducting a more detailed and thorough analysis.

\section{Key Findings}
\label{sec:key_finding}

This section lists the major key findings.
\begin{enumerate}

\item\textbf{Traditional elicitation approaches and prototyping are the most popular requirements elicitation methods for mobile app development.}

The interviewees discussed that mostly various traditional approaches like interviews, workshops, and analysis of existing solutions are used in addition to prototyping for requirement elicitation for mobile app development. One of the interviewees also stated that challenges in these approaches are usually user involvement in the project. The interviewee who mentioned this issue was developing an application for a village and conducted workshops for requirement elicitation. However, only seniors citizen attended the workshops. This caused a lack of involvement of young people in the requirement elicitation activities.

\item\textbf{A large number of interviewees rely on information provided by the crowd. They use data from app stores for analyzing existing applications for their own requirement elicitation process. In addition, some of them also consider social media for this purpose.}

The majority of the interviewees (8 out of 11) use app stores for requirement elicitation. Three of them (3 out of 8) also use social media in addition to app stores. However, the rest (5 out of 8) do not mine social media because of the required time and effort to mine additional resources in comparison to information obtained from these sources.

 Our result identified that 3 out of 11 interviewees do not consider existing applications for requirement elicitation. They stated that it is not beneficial to mine app store because of the type and domain of application. For example, one interviewee stated: 
 ``Many factors play a role not to explore app store because we are living in financial factors, we have primary restrictions and rules from legal perspectives and clients perspectives.''

\item\textbf{We identified four key elements for manual app mining: app descriptions, user feedback, screenshots, and the app itself.}\\

These four key elements are available for each application and can extract some information related to the app. Our study results suggest that app descriptions are usually not enough to gather information about an app. In addition, user reviews and screenshots are used as well. Also, applications are typically downloaded and tried for a better understanding. Another interesting parameter is watching tutorial videos about the app, but it is least used for information extraction.

\item\textbf{A large number of interviewees stated that user feedback plays a key role in the app mining process.}

The majority (8 out of 11) of the interviewees think that user feedback is a primary source to understand users of a specific feature better. The remaining ones indicated that crowd feedback is not reliable because of many fake and automatically generated reviews written for ratings and marketing purposes. They also mentioned that users cannot convey their feedback properly.

\item\textbf{Negative reviews from the crowd are also helpful and influence the feature selection processes.}

The practitioners are not only interested in positive feedback. They also search for negative reviews because it helps them to identify the market gap. They mentioned that negative reviews are helpful to build a successful application by providing a better solution to compete with similar applications.

\item\textbf{No interviewee stated to be using yet an automated tool for feature extraction from app descriptions.}

None of our interviewees are using any automated tool for feature extractions from app descriptions yet. As they perceived that app description is short and structured, it is not difficult and time-consuming to extract features manually. On the other hand, they mentioned the need for an automated tool for user feedback analysis due to the large size in order to save time and effort. They are not aware of any automated tool for user feedback analysis.

\item\textbf{The most targeted  NFRs in the app mining process are usability, user experiences and performance.}

Our results showed that practitioners also target NFRs during the analysis of existing apps. For this purpose, app descriptions are not helpful because it usually does not provide information about NFRs. The information is usually obtained from screenshots, user reviews, and application testing.

\item\textbf{Key influential factors for particular feature selections from app stores are the number of app downloads, the number of app reviews, estimated feature implementation cost, and customer/user acceptance.}

We further asked what factors influence the decision for choosing a feature from existing applications. Mostly, features are shortlisted from existing applications based on the number of app downloads and its reviews. For the final decision, feature implementation cost and customer/user acceptance are estimated.


\item \textbf{A majority of interviewees showed the need for a holistic tool for features suggestion for new apps.}

When we mentioned our proposed tool idea to extract features from app description and suggest features based on user feedback about those features (explained in detail in \ref{sec:RoadMap}), the majority showed a positive response and need for this tool. Our interviewees showed curiosity to search for feature related user responses from different sources. Most interviewees thought that such a tool would be more suitable for product based and long term projects. The main reservations related to the use of our proposed tool were completeness and correctness of the results. We also received some suggestions related to the implementation. For example, keeping the original feature list and tool generated feature list separate for keeping the originality of new applications. Another suggestion is to provide both positive and negative user reviews for features separately.

\end{enumerate}

From our key findings, we conclude that there is a need to filter user feedback according to the features of existing apps. To achieve this, practitioners are doing app store mining but currently rely on manual efforts. Our interviews corroborated that there is a need for an automated tool that suggests features for new application development using feature-wise feedback from existing apps.

\section{Early Conceptual Solution}
\label{sec:RoadMap}
\begin{figure*}[h!]
 \includegraphics[width=\linewidth]{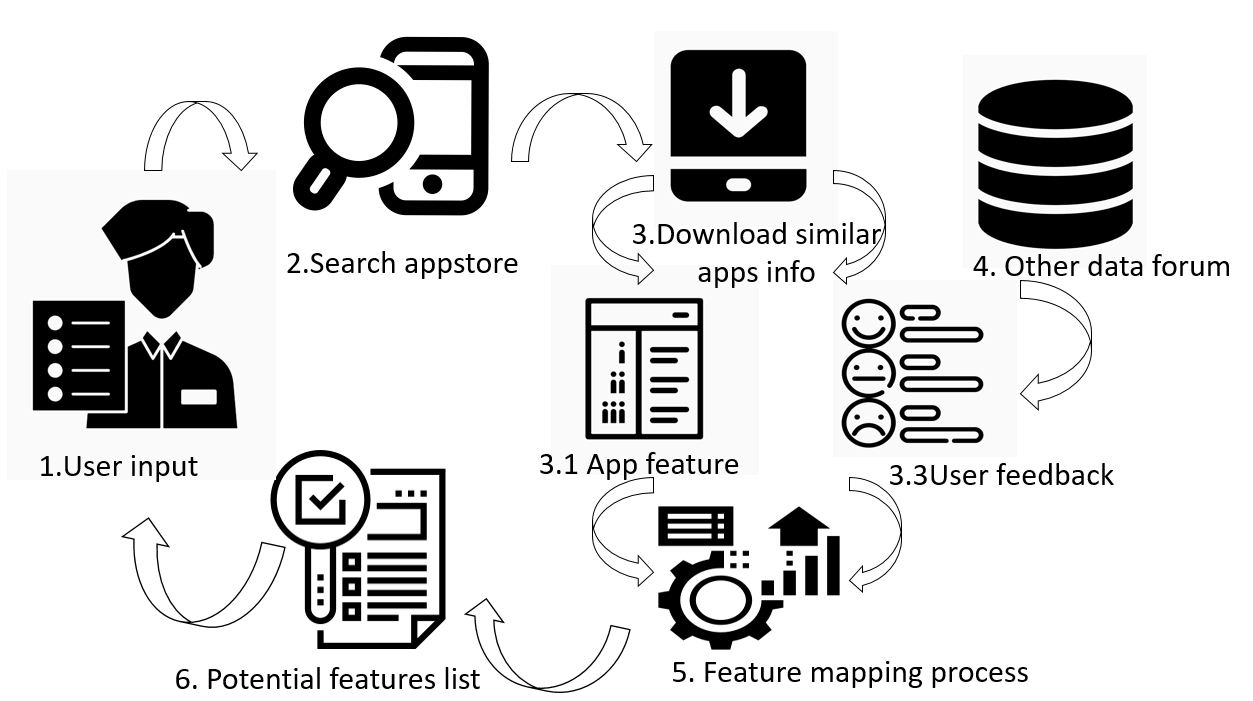}
  \caption{Simplified sketch of our envisioned solution.}
  \label{fig:methodolgy}
\end{figure*}

Our empirically study preliminary results show that practitioners are interested to have an automated tool that can help them to obtain initial requirements from existing similar applications for new applications. We depict our early vision of our proposed solution approach
in Fig.~\ref{fig:methodolgy} and explain it further in the following. 
\\
In principle, we envision the following steps:

\begin{enumerate}
    \item The proposed system takes user input in the form of the desired application category. Optionally, a user may provide the desired feature list.
    \item Based on user input app store searches similar applications in the particular domain. 
    \item Basic information of all apps in that particular category is extracted from app stores. Features for all apps(using \cite{SAFE}) and user feedback are extracted from app stores.
    \item Crowd feedback from other data forums e.g. twitter, internal wikis are extracted and connected to each feature. User reviews not related to any of the extracted features are discarded.
    \item For each application, features are then scored on the basis of various parameters such as user feedback, sentiment analysis, ratings, and cohesiveness of features.
    \item A list of features with a high score generated in step 5 for the desired application category are presented to the user as potential features. If the user provides features in step 1, candidate features are filtered further to provide only similar features to the user provided features. It's then at the discretion of the user to select among those features for app development.
\end{enumerate}

The proposed solution aims at addressing all the five research questions. RQs 1 and 2 are already addressed in this paper. As a next step to our research, we aim at addressing RQs 3-5. While we expect many challenges to arise along with the implementations (which is why we deliberately opt for Design Science research as a paradigm), we focus in the following only on selected challenges and initial ideas related to them. 

The expected challenge to address RQ 3 is the classification of user reviews manually, as practitioners stated it as a challenging task (mentioned in key finding 6). The manual classification of user feedback is tedious and difficult. To overcome this problem, one possible solution is to apply machine learning and classify user reviews into informative and non-informative categories, further performing features-wise classification of informative reviews based on the feature extracted from the app description. As a next step, we will investigate which algorithm is more suited for this classification. We will also experiment with semi-supervised algorithms to reduce labeling effort and time.

For addressing RQ 4, we have already identified some factors in our preliminary results that influence the feature selection process, i.e. our key finding 4 and 5  lead us to employ also mechanisms to conduct sentiment analyses. However, there might be other factors which we do not know yet. As the next step for RQ 4, we will, therefore, explore further factors potentially useful for the feature selection process that ultimately help to provide optimize feature list. 

Our key finding 3 gave already a good starting point on app store mining process for addressing RQ 5. For detecting similar features from existing apps (RQ5), one possible solution is to use a clustering algorithm. We will investigate which clustering or other possible machine learning algorithm might work efficiently to map similar features.

Regardless of the specific research question, we are certainly aware that there exists much work related to ours on which we can build our holistic approach. The contributions by Harman et al. \cite{Harman2012AppSM} and by Johann et al. \cite{SAFE}, for instance, extract features from app description and the work by Guzman et al. \cite{Guzman} and by Gu et al. \cite{X_Gu} analyze feature feedback analysis. Further, the study by Martin et al. \cite{7180073} to categorize similar application is similar to our envisioned steps. We will investigate to what extent existing solutions could become part of our tool-supported approach. 

\section{Related work}
\label{sec:Related_Work}

The research work presented in this section is not extensive, and only lists literature closely relevant to our research. An overview on the automation of various tasks in the overall requirement engineering phase is presented in \cite{Tahira}. The current research, benefits, and challenges in crowd based requirement engineering are presented by Groen et al. \cite{Eduard.C}. The existing literature classifies user feedback either into bug reports and new feature requests, or into functional and non-functional requirements \cite{{Maleej2015},{Lu:2017},{Deocadez:2017},{Jiang:2015}}. There is also a promising trend to mine social media forums, such as Twitter, for extracting users requirements for software evolution \cite{{NayebiRE},{Nayebi},{Guzman},{Williams}}. The idea to reduce domain analysis efforts for developing applications has been explored in \cite {Cleland:2011}, \cite{Nayebi_superMobileApp:2017},and 
\cite{Nayebi_Mcmurray:2017}.

A recommender system is proposed by Dumitru et al. and Hariri et al. in \cite{Cleland:2011} and \cite{Cleland:2013} respectively. This recommender system mines product descriptions from a publicly available online repository, Softpedia. It initially takes product description from the user, mines repositories on that basis, and then provides recommended feature candidates to be included in the user intended software system. Text mining and incremental diffusive clustering algorithm are further used for domain-specific feature identification. Afterward, kNN (k-Nearest Neighbor) algorithms are applied for product-specific feature recommendations. 

Nayebi et al. proposed an approach to mine existing application descriptions from app stores and set their primary use case for release planning \cite{Nayebi_superMobileApp:2017}. It uses the algorithm mentioned in \cite{Harman2012AppSM} for feature extractions from app descriptions. Extracted features are optimized on the basis of the estimated values of features and cohesiveness between features. A new model using bi-criterion integer programming is proposed to solve this optimization problem.

Guzman et al. further proposed a method to identify user preferences for specific features by mining user reviews \cite{Guzman_2014}. It extracts features from the user feedback available on app store based on the frequency of occurrence of words, and then determines user preferences by analyzing the sentiments in user reviews.
  
Another study by Nayebi et al. proposed a method named as MAPFEAT, extracting user needs from Twitter and mapping them to features of already existing apps \cite{Nayebi_Mcmurray:2017}. This technique particularly mines Twitter for event-based tweets, and extract topics out of them. It then searches all the applications in app stores for features based on these topics in conjunction with crowdsourcing, identifying missing features in currently existing apps.   

In our opinion, these studies provide a great starting point for our research and further strengthen our confidence in the importance of work in this area. However, we consider extending some approaches not only content-wise, but also with respect to their field of application. For instance, in \cite{Cleland:2011} and \cite{Cleland:2013}, features are extracted from an online repository with no crowd user feedback support. Due to the similarities with our idea, we are considering to extend that work to platforms allowing for user feedback. Although \cite{Nayebi_superMobileApp:2017} used app stores to perform feature extractions as we intend to do, their solution does not consider user feedback and only takes app descriptions. Guzman et al. \cite{Guzman_2014} relies on app stores and user feedback as well, but they do not suggest features for similar new applications development; the features not mentioned in the app description are not filtered out as well. The idea of searching app stores for feature suggestions to evolve existing apps is used in \cite{Nayebi_Mcmurray:2017} as well, but app stores are not searched on the basis of categories, resulting in generalized features suggestions. Also, all such analysis solutions focus on analyzing user reviews without linking them to descriptions of the corresponding app.

Compared to the existing publication landscape, we are not only relying on app description, and extend this potential requirements source by further considering features related feedback. Our solution intends to support the classification of feedback based on the extracted features from the app description. This shall help to analyze users’ feedback on specific features. We will also be integrating feedback from the other sources, i.e from social media or internal organizational feedback. This integration brings a challenging task in order to build up a general model to deal with different types of data and their associated metadata. Finally, we will map and suggest features from existing apps to support the development of new apps and, thus, extend current focus areas of software evolution to the design and implementation of new apps.


Similar to our interview study, empirical studies have been conducted as well in \cite{Nayebi_Survey} and \cite{Pagano:2013}. These studies focus on requirements engineering practices for software release planning of existing applications. Another interview and questionnaire-based approach discussed in \cite{F_Sarro:2018} elaborates current software engineering practices in application development. This study covers practices for an overall software life cycle including requirements, testing, and maintenance. One further delineation from existing work is that we focus on current industry practices for new applications that have not yet been developed. Our research goal is more narrowed and focused on that scope, i.e. we intend to particularly support and improve requirement engineering practices for developing new applications.

\section{Conclusion}
\label{sec:conclusion}

The app store is full of information provided by the crowd, and mining the app store has already shown promising results to support software and requirements engineering, in particular in the context of software evolution. This paper proposes a tool-supported approach that analyzes apps and associated feedback on the feature level. This information is used to inspire the development of new apps and in particular to suggest features for developing similar new apps. We derived five research question for implementing this proposed approach. To this end, we conducted an interview study for addressing the first two research questions. We identified that practitioners currently are mining the app store manually, and showed the need for an automated tool solution. As future work, we will address the remaining research questions on the basis of our preliminary results and will develop the proposed automated tool solution. As proposed in our early conceptual solution, we will therefore apply machine learning to analyze user feedback from multiple sources. We plan to build a general model for automating requirements elicitation and mining techniques considering different types of data and their associated metadata. We cordially invite researchers to join this endeavor to further increase the efficiency of requirements elicitation practices in the future.

\bibliographystyle{abbrv} 
\bibliography{./req}  
\end{document}